# Realization of MIMO Channel Model for Spatial Diversity with Capacity and SNR Multiplexing Gains

**Subrato Bharati[1], Prajoy Podder[2], Niketa Gandhi[3] and Ajith Abraham[4]**

[1] Department of EEE, Ranada Prasad Shaha University, Narayanganj-1400, Bangladesh
*subratobharati1@gmail.com*

[1,2] Institute of Information and Communication Technology,
Bangladesh University of Engineering and Technology, Dhaka-1000, Bangladesh
*prajoypodder@gmail.com*

[3] University of Mumbai, Maharashtra, India
*niketa@gmail.com*

[4] Machine Intelligence Research (MIR) Labs, Auburn, Washington, USA
*ajith.abraham@ieee.org*

*Abstract*: **Multiple input multiple output (MIMO) system transmission is a popular diversity technique to improve the reliability of a communication system where transmitter, communication channel and receiver are the important elements. Data transmission reliability can be ensured when the bit error rate is very low. Normally, multiple antenna elements are used at both the transmitting and receiving section in MIMO Systems. MIMO system utilizes antenna diversity or spatial diversity coding system in wireless channels because wireless channels severely suffer from multipath fading in which the transmitted signal is reflected along various multiple paths before reaching to the destination or receiving section. Overwhelmingly, diversity coding drives multiple copies through multiple transmitting antennas (if one of the transmitting antenna becomes unsuccessful to receive, other antennas are used in order to decode the data) for improving the reliability of the data reception. In this paper, the MIMO channel model has been illustrated. Moreover, the vector for transmitting signal has been considered by implementing least square minimization as well as linear minimum mean square estimation. Parallel transmission of MIMO system has also been implemented where both the real part and imaginary part of the original, detected and the corresponding received data sequence has been described graphically. One of the important qualities of MIMO is a substantial increase in the capacity of communication channel that immediately translates to comparatively higher signal throughputs. The MIMO communication channels have a limited higher capacity considering the distortions for various deterministic channel recognitions and SNR. The MIMO channel average capacity is achieved more than 80% for dissimilar levels of impairments in transceiver when the value of kappa (Level of impairments in transmitter hardware) reduces from 0.02 to 0.005. The finite-SNR multiplexing gain (Proportion of MIMO system capacity to SISO system capacity) has been observed for deterministic and uncorrelated Rayleigh fading channels correspondingly. The core difference is in the high SNR level. It may occur for two reasons: (a) there is a quicker convergence to the limits under transceiver impairments (b) deterministic channels that are built on digital architectural plans or topographical maps of the propagation environment acquire an asymptotic gain superior than multiplexing gain when the number of transmitting antenna is greater than the number of receiving antenna.**

## I. Introduction

In the last few years, MIMO has developed as an embryonic technology to deal with substantial capacity for mobility as well as high data rates required by the wireless communication systems for next generation. By means of multiple transmit along with receive antennas, a MIMO scheme achieves spatial diversity, advanced data rate, better coverage besides established link robustness deprived of snowballing overall transmission bandwidth or power [1-4]. On the other hand, MIMO depends on the information of Channel State Information (CSI) at data decoding as well as detection for the receiver [5]. It has been said that while the channel of the network is flat fading besides impeccably recognized to the receiver, the presentation of a MIMO scheme develops linearly with the quantity of receive otherwise transmit antennas each less. Consequently, a robust, as well as accurate channel approximation, is of vital prominence for intelligible demodulation in MIMO schemes which is wireless [6, 33, 34]. The custom of MIMO frequencies and channel, when bandwidth is inadequate, has considerably greater spectral efficiency as opposed to Single-Input Single-Output (SISO), Multi-Input Single-Output (MISO) as well as Single-Input Multi-Output (SIMO) channels [7]. It is exposed that the extreme realizable diversity gains of MIMO channels, that the invention of the number of the receiver as well as transmitter antennas. For that reason, by using the channels of MIMO not only the flexibility of wireless communications can be amplified, but also its toughness [8]. Mobile communication schemes transmit data by varying the amplitude otherwise phase of the waves of radio. In the unloading side of the mobile scheme, amplitude or else phase can change extensively [9]. On the other hand, these indicators oblige information on the channel impulse response (CIR), which can be delivered by a discrete channel predictor to diminish the probability of error [10].

Recently, lot of research has been deliberated MIMO communications incited by the high SNR scheme for stimulating capacity scaling [11]. Emil et al. [11] described the applicability in cellular complexes and uncertain expansions of MIMO network in excess of conventional systems, besides the output might even decrease owing to the additional overhead. One description is the limited channel soundness time that bounds the properties for channel achievement as well as coordination among nodes, therefore constructing a limited essential ceiling on behalf of the spectral efficiency in the power of networks irrespectively in addition to the quantity of antennas.

The contribution of this paper can be summarized as follows: A MIMO system is proposed with spatial diversity and high-speed transmission where capacity and SNR multiplexing gains is considered under different conditions. LSE, LMMSE with visualization of a real part and imaginary part of the sequence of the signal are also illustrated in this paper. Transmitter hardware impairments level (value of kappa=0.4) according to high SNR limit for ideal hardware provides higher value (6.7926). On the contrary, the value of kappa 0.01 provides lower capacity and SNR value than others. A capacity limit is [15.4373, 18.7744] for kappa value of 0.4. The algorithm of this new method is described and compared it with the literature. Section 3 defines MIMO channel describing LSE and LMMSE. Power distribution for water-fill algorithm is illustrated in Section 4. Section 5 and 6 also describes capacity limits of MIMO channel and simulation results respectively.

## II. Literature Reviews

Björnson [11] stated that the ideal MIMO model capacity has an extraordinary SNR grade generating the lowest of the quantity of transfer and receiver antennas. Their result showed that they have finite level SNR. It has been showed systematically that the proposed MIMO channels have a limited upper limit of capacity, for slightly distribution of channel beside SNR. The SNR value thus falls to zero (0). But the comparative gain of the capacity of using MIMO for different kappa value has been focused in this research work. The improvement of this research work has been compared with the work of Emil Bjornson. The average capacity is increasing with the decreasing of kappa value at a point, capacity is not increased. The proposed research work described better capacity than other research [11,15].

Chaaban et al. [12] illustrated intensity modulation traditional appreciation of MIMO scheme. Lower limits of the capacity are recognized in their paper by evolving the achievable proportions of two allowed schemes. Their technique is optical related. The proposed wireless communication method is antenna related for a different number of MIMO antenna. Besides, their capacity higher limits are consequential and associated with the minor limits. Consequently, the capacity of high SNR is reflected for the condition where the transmit spaces quantity is not higher than the receive spaces quantity, then is showed to be achievable by the decomposition in QR scheme. We also provided high SNR capacity for MIMO system with capacity limits. Their proposed high SNR capacity and capacity limit differentiated as the proposed new method. Because their method is optical based but in this paper method is wireless MIMO system base where detect LMMSE and also implement high SNR capacity for different transmitter hardware impairments level (kappa). Ge et al. [13] proposed a massive MIMO system for 5G system. It can expressively improve the scheme with high spectrum capacity and efficiency of energy. Massive MIMO schemes, estimation in accurate channels are an interesting problem, particularly when the quantity of considerations to be appraised is enormous then the quantity of pilots is bounded. In their paper, a compression-based estimation of LMMSE channel approximation algorithm is offered for 5G massive MIMO systems. But this research work also designated MIMO scheme for cellular network. This research paper is capable of detecting the sequence of LMMSE with power distribution by waterfill algorithm and improving the high SNR capacity.

Liu et al. [14] presented the corresponding conditions among the decoders besides detector for MIMO-NOMA scheme, which is previously used to improve the achievable rate of the recognition of iterative LMMSE. They verify that a complemented detector of iterative LMMSE can accomplish the ideal capacity of MIMO-NOMA symmetric with some amount of operators, the ideal sum capacity asymmetric of MIMO-NOMA by way of some amount of operators, all the greatest exciting opinions in the capability area of MIMO-NOMA with some amount of operators [14]. In this paper detected a sequence of LSE, LMMSE for MIMO system with multiple antenna and provided high average SNR capacity.

E. Björnson et al. [15] proposed another method which detected unlimited capacity for a massive MIMO system. The cellular networks capacity can be developed by the exceptional array gain besides spatial multiplexing present by Massive MIMO. As its initiation, the intelligible interference affected by pilot contagion has been understood to generate a limited capacity bound, as the quantity of antennas vigor to infinity. In that paper, they show that that is inappropriate as well as an object from via models simplistic channel then suboptimal linking/precoding systems. They also illustrated that using multi-cell MMSE linking/ precoding besides a small amount of channel in spatial large-scale fading or correlation differences done by the array, the capacity growths deprived of limits as the quantity of antennas rises, level under pilot contagion. But the offered work is illustrated as the limitation channel capacity. This is the main difference in the suggested scheme.

Yoo et al. [16] offered a $2 \times 2$ MIMO scheme with virtual meta external antennas as the transfer and receiver antennas going at 5.9 GHz. They prove that the beamforming flexibility sustained by the meta external antennas can be castoff to realize the correlation of low spatial then clustered for improving SNR-gain in MIMO channels, prominent to an important development of the capacity in the channel. Mathematical studies display 2.11-fold, 2.36-fold capacity improvements in MIMO channels by two and one clusters, correspondingly, associated with a MIMO scheme containing true dipoles.

Chen et al. [17] delivered a methodical evaluation of the reciprocal pairing in multiple-input multiple output (MIMO) schemes, comprising belongings on presentations of MIMO schemes besides several decoupling methods. The reciprocal pairing modifications the antenna features in an array as well as, consequently, destroy the scheme presentation of the MIMO scheme and bases spectral regrowth. Though the scheme presentation can be incompletely better-quality by normalizing out the reciprocal pairing in the domain of digital, it is added operative to custom decoupling methods (starting the antenna point) to incredulous the reciprocal pairing properties. Nearly standard decoupling methods for MIMO schemes are also offered. It is exposed that reciprocal pairing below -10 dB has an insignificant consequence on error rate or the capacity presentation of the MIMO scheme. However, when allowing for the nonlinearity in PA, the out of band emission can be compact by decreasing the reciprocal pairing (level for reciprocal pairing below -28 dB).

Sahu et al. [18] proposed the channel estimation agreed to the receiver to recompense for differences in the message channel so that decrease the error probability in the detected signal. The estimate is developed by transfer reference signals recognized as pilots which are known at the receiver. The BER of MMSE evaluation has about gain 5-15 dB in SNR beyond the LS approximation for the comparable rates of BER. The main disadvantage of the MMSE estimation is its extraordinary complexity. The proposed method detected LMMSE then described SNR capacity gain. The paper also detected LE and LSE data sequence.

Obinna Okoyeigbo [19] determined the characteristics of the channel besides how it touches transferred signals are recognized as the estimation of the channel. MIMO-OFDM is a mixture of OFDM and MIMO methods. In their paper, relative performance of SISO, MISO and MIMO channel approximation, via OFDM was implemented. The MMSE is resilient to noise, then overtakes the LS, thru 10-15dB SNR gain. The LS involves an improving SNR to competition the performance of MMSEs. But this research paper has been designated the performance of LMMSE besides recognition with high SNR capacity in decibel scale. Where the highest capacity limits [67.7543, 71.4389].

## III. MIMO Channel

The utilization of numerous antennas at the transmitter side in addition to receiver side (Multiple-input multiple-output (MIMO) can effect a substantial increase of the capacity. This is owing to two special effects: (i) diversity, such as toughness against the channel of fading among a transmitter as well as a receiving antenna, and (ii) space-time coding, such as the transmission is parallel of the information thru several transmit antennas.

On the other hand, this increase of capacity was constructed on a significant statement: all channels among the transmit antennas as well as the receive antennas are exactly recognized [20-21, 35]. In transmitter section, the quantity of antennas is symbolized by T moreover the comparable number in the receiver section is indicated as R. If R=T, it is recycled for high transmitting data. If R>T, we realize mutually spatial diversity as well as transmission for high speed (this is generally castoff).

| Authors Name | Year | Adopted Method | Core Findings |
|---|---|---|---|
| E. Björnson et al. [11] | 2013 | Implementation of MIMO channel | Multiplexing Gains with capacity limit |
| Chaaban et al. [12] | 2018 | Intensity-Modulation of MIMO in optical channels | Capacity of high-SNR with Capacity Bounds |
| Ge et al. [13] | 2019 | 5G systems of massive MIMO in adaptive sparsity | Compression based estimation of LMMSE channel |
| Liu et al. [14] | 2019 | Capacity-Achieving in MIMO-NOMA | LMMSE Detection |
| Emil et al. [15] | 2018 | Massive MIMO | Unlimited Capacity |
| Yoo et al. [16] | 2019 | MIMO Channels Clustered | Capacity of Spatial Multiplexing |
| Chen et al. [17] | 2018 | MIMO system | Mutual Coupling |
| Sahu et al. [18] | 2014 | Estimation method for MIMO-OFDM | MMSE and LS |
| Okoyeigbo et al. [19] | 2018 | MMSE and LS of SISO, MIMO, MISO | Relative performance of MIMO-OFDM |

*Table 1.* Summary of the literature review

The link among the antenna t in the section of transmitter besides the antenna r in the section of receiver is represented as the channel of flat fading. Therefore, the model is symbolized as $= G * b + n$. Here G is the Gaussian channel matrix (such as base band-< pass band-< base band using the theory that where flat in this channel), the vector of transmitter is a, the vector of receiver is b and the noise vector is n. The vector of transmitter a is expected via the following method:

1. By estimating least square minimization (LSM) (formerly known as zero-forcing error), this paper has been evaluated by minimizing ‖b-Ga‖². The explanation is acquired as

$$\hat{a} = (G^T G)^{-1} G^T b \qquad (1)$$

2. By estimating linear minimum mean square estimation(LMMSE), wherever $E(||\hat{a} - a||^2)$ is reduced with $\hat{a} = C^H b$ (with linear approximation), it is initiate that the peak rate of C is achieved as

$$C = E(bb^H)^{-1} E(a^H b) \qquad (2)$$

3. $\sigma_n^2 I$ is covariance matrix of noise vector where $P_D I$ is the covariance matrix and I is the identity matrix, the probable vector is acquired as

$$\hat{a} = P_D G^H (P_D G G^H + \sigma_n^2 I)^{-1} b$$
$$= P_D (P_D G^H G + \sigma_n^2 I)^{-1} G^H \qquad (3)$$

The LMMSE is calculated via $P_D$ (the signal power with its average). This peak rate is the estimation uncertainty the provisional subsequent probability concentration function of $\hat{a}$ specified $\hat{b}$ is Gaussian as well as the estimation declared now is the provisional mean, provisional median otherwise provisional mode of the subsequent possibility concentration function.

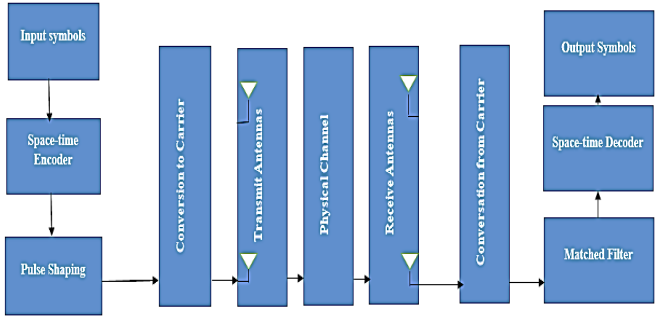

**Figure 1.** Block diagram of MIMO Communication Scheme

---

### Algorithm 1: MIMO Channel

---

i) Channel matrix input

$$C = \begin{bmatrix} 0.5 + 0.6 * i & 0.15 + 0.85 * i & 0.4 + 0.10 * i \\ 0.8 + 0.5 * i & 0.81 + 0.86 * i & 0.05 + 0.87 * i \end{bmatrix}^t$$

ii) Transmitted sequence, $T = [T_1 \quad T_2]$
$$P = C * T + \sqrt{0.01} * rand(3,100) + i * \sqrt{0.01} * rand(3,100)$$

iii) Detection of receiver using LMSE
$$R_{LS} = (C^t * C) * C^t * P;$$

iv) Implement R1_LS and R2_LS;

v) Set Ry1 and Ry2 to a suitable value

    for k=1:1:200

        $R_{y1} = R_{y1} + Y(:,k) * Y(:,k)^t;$

        $R_{y2} = R_{y2} + Y(:,k) * T(:,k)^t;$

    end

        $R_{y1} = (\frac{1}{99}) * R_{y1};$

        $R_{y2} = (\frac{1}{99}) * R_{y2};$

        $LE = (R_{y1})^{-1} * R_{y2};$

vi) LE estimation= $(LE)^t * Y;$

---

#### A. Transmitted data arrangement in the MIMO system

The impulse response is represented by $g_P(\theta_R, \varphi_R, t, \tau, \theta_T, \varphi_T)$ which is connected field released. That is transmitted array to the incident of field on the obtain array. The requirement on time (t) proposes that this response of impulse is the variant of time due to motion of scatters in the transmission motion or environment of the receiver as well as transmitter. The variable τ denotes the delay of time comparative to the period of excitation $t$. We adopt an impulse response which is finite, so that $g_P(\theta_R, \varphi_R, t, \tau, \theta_T, \varphi_T) = 0$ on behalf of $\tau > \tau_0$. We moreover adopt that $g_P(\theta_R, \varphi_R, t, \tau, \theta_T, \varphi_T)$ residues perpetual over the interval of period *(in t)* of "$\tau_0$" duration to facilitate over a distinct transmission, the channel can be preserved when it was physical as a linear, time-invariant scheme. Adopt now that $a_X(t)$ is the input signal which generates the field $a_P(\varphi_T, t, \theta_T)$ emitted as of the transmit array, wherever the elevation is denoted by $(\theta_T, \varphi_T)$ besides the angles of azimuthal occupied with due to the coordinate structure [22]. At the obtain array, the distribution of field $b_P(\varphi_R, t, \theta_R)$, here angles referenced is denoted by $(\varphi_R, \theta_R)$ to the obtain array coordinate structure, then can be represented as the convolution

$$b_P(\theta_R, \varphi_R, t) = \int_0^{2\pi} \int_0^{\pi} \int_{-\infty}^{\infty} g_P(\theta_R, \varphi_R, t, \tau, \theta_T, \varphi_T) b_P(t - \tau, \theta_T, \varphi_T) cos\theta_T d\tau d\theta_T d\varphi_T$$

(4)

The explicit description of this channel can reliant on the aim of the investigation. Perhaps, in some circumstances we desire to concentrate on the somatic channel which response is impulse as well as use it to produce a matrix of channel connecting the signal $a_X(t)$ then $b_X(t)$. A mutual statement will be that totally smattering in the transmission channel is in the remote field besides that a separate quantity of transmission "paths" links the transmitter as well as receiver arrays [23]. In this theory, the response of physical channel for L pathways may be described as

$$g_P(\theta_R, \varphi_R, t, \tau, \theta_T, \varphi_T) = \sum_{l=1}^{L} X_l \delta(\tau - \tau_l) \delta(\theta_T - \theta_{T,l}) \times \delta(\theta_R - \theta_{R,l}) \delta(\varphi_T - \varphi_{T,l}) \delta(\varphi_R - \varphi_{R,l})$$

(5)

Here the gain is $X_l$ is the $l$ th pathway with departure angle $(\varphi_T, \theta_T)$, arrival angle $(\varphi_R, \theta_R)$, then arrival time τ. Dirac delta function is represented in terms of $\delta(\cdot)$. The variation of time of the channel is involved by creation the multi paths limits of $(L, X_l, \tau_l, \theta_{T,l})$ time reliant. To custom this reaction to transmit $a_X(t)$ to $b_X(t)$, it is at ease to continue in the domain of frequency. For applicable signals, We can get the Fourier transform to achieve $\widetilde{a_X}(\omega), \widetilde{b_X}(\omega), \widetilde{\eta}(\omega)$ then get the Fourier transform of $g_P$ according to the variable of delay τ to achieve $\widetilde{g_P}(\theta_R, \varphi_R, t, \tau, \theta_T, \varphi_T)$ where the radian frequency is ω. $\beta_l$ is channel transfer function, We essentially convolve these outlines with $\widetilde{g_P}$ in the angular coordinates to achieve

$$\widetilde{b_X}(\omega) = G_P(\omega) \widetilde{b_X}(\omega) + \widetilde{\eta}(\omega)$$

(6)

Where,

$$G_{p,mn}(\omega) = \sum_{l=0}^{L} \beta_l \varepsilon_{R,m}(\theta_{R,l}, \varphi_R, \omega, l) \varepsilon_{T,n}(\theta_{T,l}, \varphi_T, \omega, l)$$

(7)

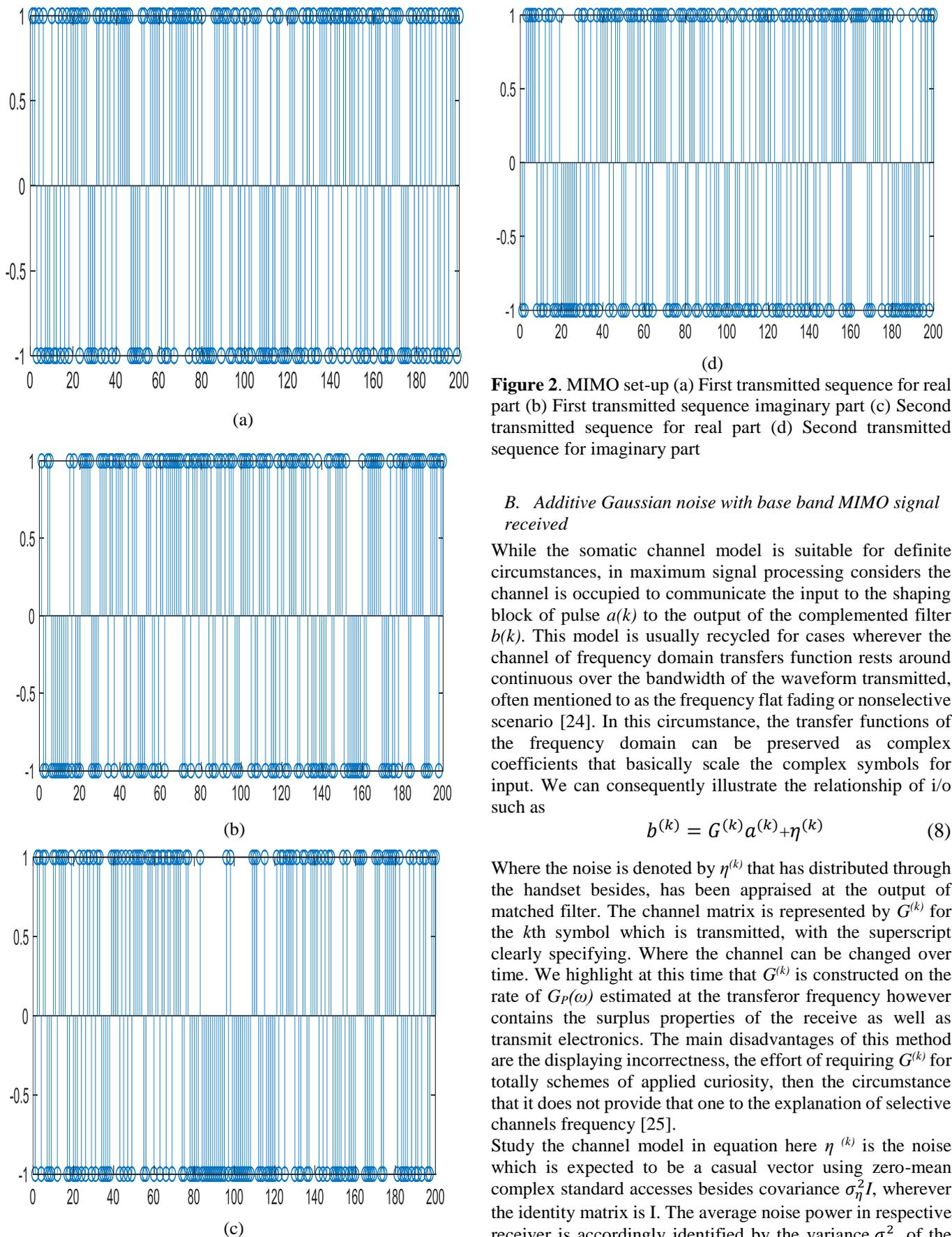

**Figure 2**. MIMO set-up (a) First transmitted sequence for real part (b) First transmitted sequence imaginary part (c) Second transmitted sequence for real part (d) Second transmitted sequence for imaginary part

### B. Additive Gaussian noise with base band MIMO signal received

While the somatic channel model is suitable for definite circumstances, in maximum signal processing considers the channel is occupied to communicate the input to the shaping block of pulse $a(k)$ to the output of the complemented filter $b(k)$. This model is usually recycled for cases wherever the channel of frequency domain transfers function rests around continuous over the bandwidth of the waveform transmitted, often mentioned to as the frequency flat fading or nonselective scenario [24]. In this circumstance, the transfer functions of the frequency domain can be preserved as complex coefficients that basically scale the complex symbols for input. We can consequently illustrate the relationship of i/o such as

$$b^{(k)} = G^{(k)}a^{(k)} + \eta^{(k)} \qquad (8)$$

Where the noise is denoted by $\eta^{(k)}$ that has distributed through the handset besides, has been appraised at the output of matched filter. The channel matrix is represented by $G^{(k)}$ for the $k$th symbol which is transmitted, with the superscript clearly specifying. Where the channel can be changed over time. We highlight at this time that $G^{(k)}$ is constructed on the rate of $G_P(\omega)$ estimated at the transferor frequency however contains the surplus properties of the receive as well as transmit electronics. The main disadvantages of this method are the displaying incorrectness, the effort of requiring $G^{(k)}$ for totally schemes of applied curiosity, then the circumstance that it does not provide that one to the explanation of selective channels frequency [25].

Study the channel model in equation here $\eta^{(k)}$ is the noise which is expected to be a casual vector using zero-mean complex standard accesses besides covariance $\sigma_\eta^2 I$, wherever the identity matrix is I. The average noise power in respective receiver is accordingly identified by the variance $\sigma_\eta^2$ of the problematic random variables. The average period of signal power along with totally receive ports is known as

$$P_R = \frac{1}{N_R} E\{a^{(k)G} G^{(k)G} G^{(k)} a^{(k)}\}$$
$$= \frac{1}{N_R} T_r[E\{a^{(k)} a^{(k)G}\} E\{G^{(k)G} G^{*(k)}\}]$$

$$(9)$$

Where,

$$P_R = \frac{P_t}{K N_T N_R} \sum_{k=1}^{K} ||G^{(k)}||_F^2 \qquad (10)$$

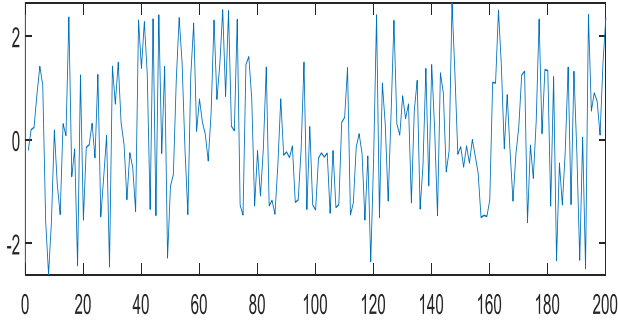

(a)

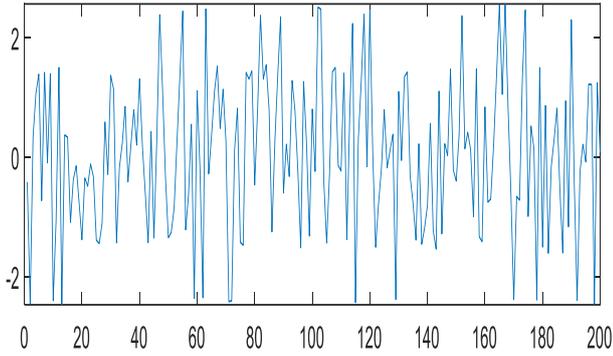

(b)

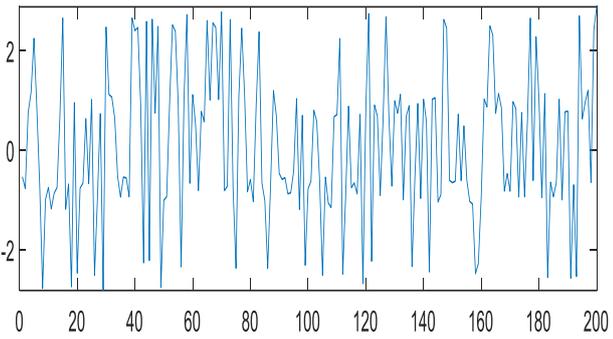

(c)

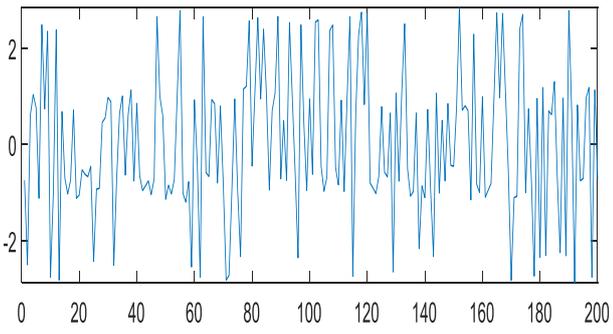

(d)

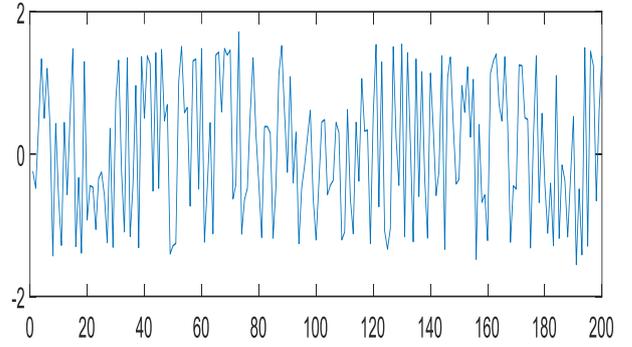

(e)

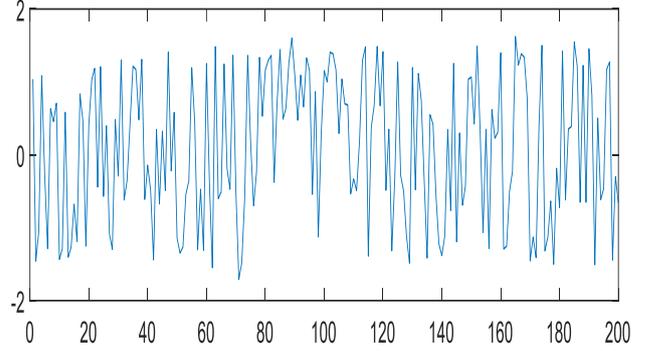

(f)

**Figure 3.** Additive Gaussian noise (a) First received sequence for real part (b) First transmitted sequence imaginary part (c) Second received sequence for real part (d) Second received sequence for imaginary part (e) Third received sequence for real part (f) Third received sequence for imaginary part

### C. MIMO System Least Square Estimation

The number of antennas in the transmitter section is assumed as M and the number of antennas in the receiver section is assumed as N. If N =M, it is used for high data transmission. If N > M, not only spatial diversity but also high speed transmission is achieved [26]. Now the receiver, the combination of a linear of completely the transferred sub-blocks are diminished via time-varying Rayleigh fading otherwise inter symbol interloping (ISI) is experiential in (AWGN), as displayed in Fig. In this paper, deprived of loss of simplification, the channel of flat-fading MIMO with Rayleigh scattering under Markov model that is provided first-order variant is implicit [26].The recognizable signal $a_k^i$ after the receiver ($i=1,...,N$) at distinct index of time $k$ is

$$a_k^i = \sum_{j=1}^{M} g_k^{i,j} s_k^j + w_k^i$$

$$(11)$$

Where the transferred symbol is $s_k^j$ from $j$-th antenna in index of time k. The propagation attenuation is $g_k^{i,j}$ between $j$-th input. In this estimation *MN* channel is cruelly fluctuate in time of data block with the subsequent autocorrelation,

$$E[g_k^{i,j}(g_l^{i,j})] \cong J_0(2\pi f_D^{i,j} T|k-l|)$$

$$(12)$$

Consistent with autocorrelation model, the normalized spectrum for each tap $g_k^{i,j}$ is,

$$S_k(f) = \begin{cases} \dfrac{1}{\pi f_D^{i,j}\tau\sqrt{1-(\frac{f}{f_D^{i,j}})^2}} & , f_D^{i,j} > |f| \\ 0, & if\ not \end{cases} \qquad (13)$$

Where,

$$g_k^{i,j} = \sum_{l=1}^{L} \alpha_{i,j,l} g_{k-l}^{i,j} + v_{i,j,k} \qquad (14)$$

Where, $g_k^{i,j}$ denotes the propagation attenuation between the j-th input and the i-th output of the MIMO channel that is a complex number with Rayleigh-distributed envelope. $\alpha_{i,j,l}$ is the l-th coefficient between the jth transmitter and the i-th receiver where $v_{i,j,k}$ are zero-mean independent identically distributed complex Gaussian processes with variances given by,

$$E\{v_{i,j,k}(v_{i,j,k})\} = \sigma_{v_{i,j,k}}^2 \qquad (15)$$

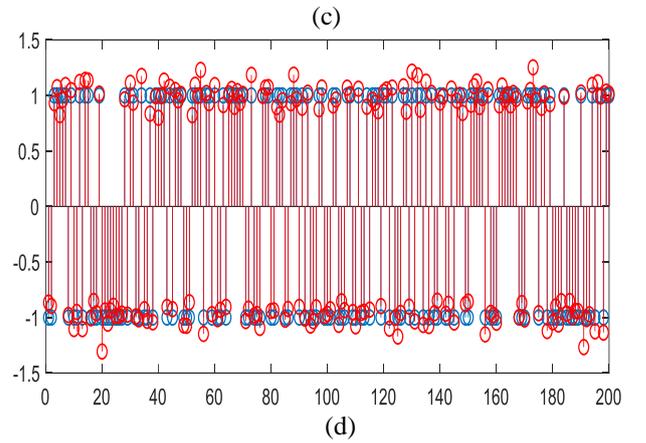

(c)

**Figure 4.** MIMO System LSE (a) First LSE sequence for real part (b) First LSE sequence for imaginary part (c) Second LSE sequence for real part (d) Second LSE sequence for imaginary part

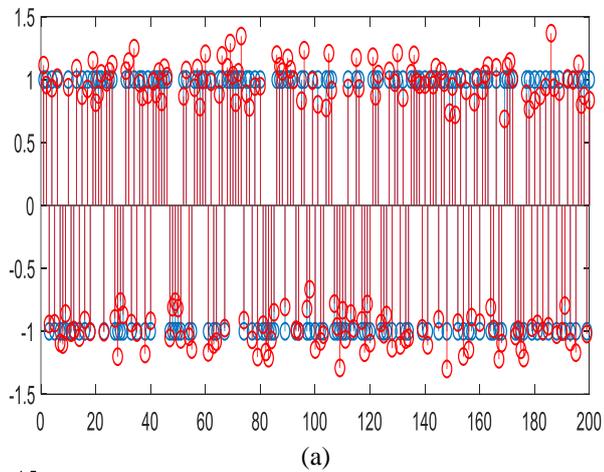

(a)

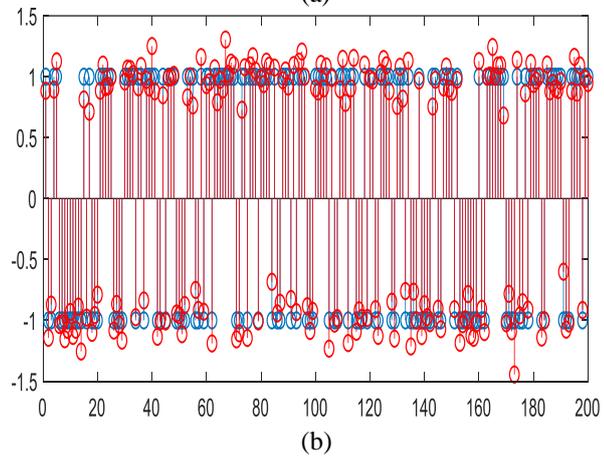

(b)

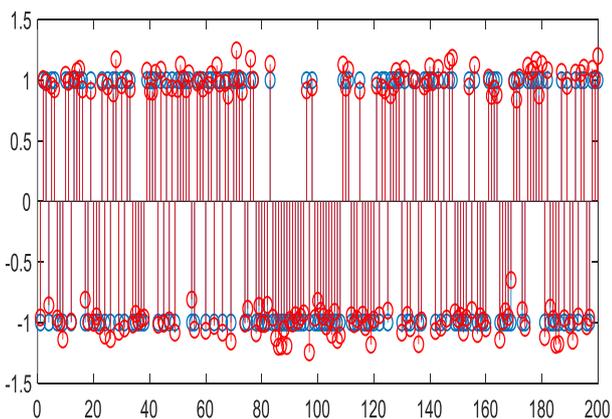

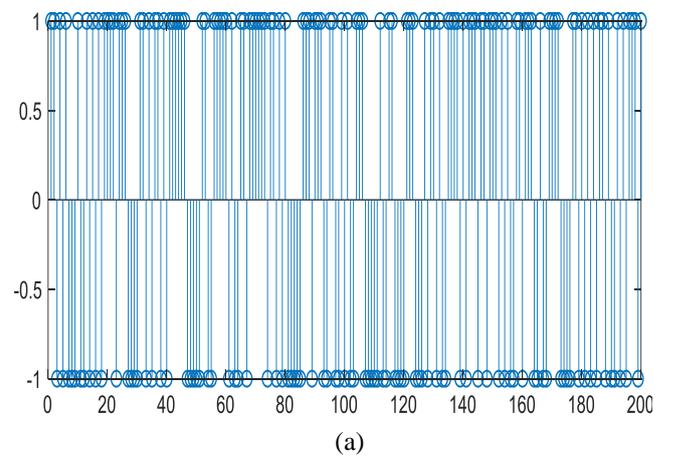

(a)

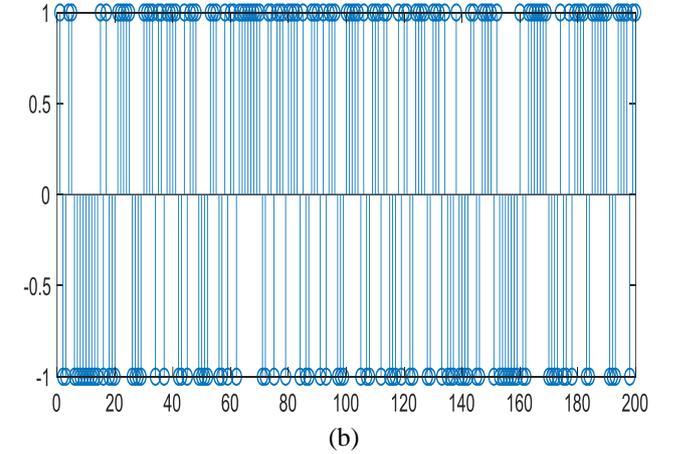

(b)

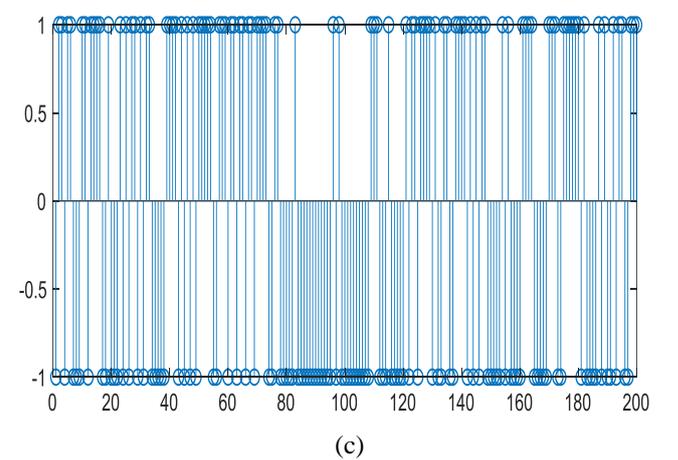

(c)

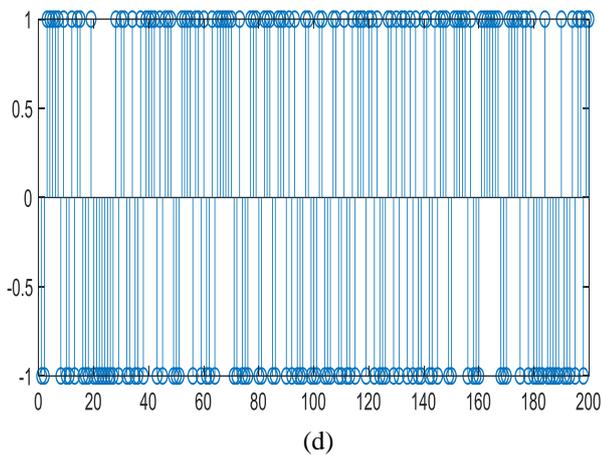

(d)

**Figure 5.** MIMO System detected sequence of LSE (a) First deceted LSE sequence for real part (b) First deceted LSE sequence for imaginary part (c) Second deceted LSE sequence for real part (d) Second deceted LSE sequence for imaginary part

### D. Linear minimum mean square estimation (LMMSE)

Let a single point to point MIMO scheme with $N_t$ and $N_r$ antennas can be assumed where $(N_r \geq N_t)$. We emphasis on the system of spatial multiplexing wherever de-multiplexed of its data stream is openly into $N_t$ steams. The overall transfer power is scattered similarly over the transfer antennas with the variance of component [27]. The scheme model can be denoted as,

$$b = G * a + n \tag{16}$$

Wherever $b \in C^{N_r \times 1}$ provides the complex vector of receiver, $a = [a_1, \ldots \ldots, a_{N_t}]^T \in C^{N_t \times 1}$ is the complex vector of transferring with $E[|a_i|^2] = \frac{E_s}{N_t}$, $G \in C^{N_r \times N_t}$ is the fading matrix of complex Gaussian with component variance, and $n \in C^{N_r \times 1}$) is self-regulating identically scattered AWGN using zero mean then variance $N_0$.

$$h_i = (GG^G + \sigma^2 I)^{-1} g_i \tag{17}$$

Where $\sigma^2 = \frac{N_t N_0}{E_s} = \frac{N_t}{Signal\ to\ Noise\ Ratio}$, $g_i$ is the column ($i$-th) of G, besides identity matrix is $N_t \times N_r$. Put on filter vector keen on equation (16) and we get

$$z_i := H_i^G b = \beta_i a_i + w_i \tag{18}$$

Wherever $\beta_i = h_i^G G_i$ then the interfering plus noise duration $w_i$ is illustrated as $\sum_{i \neq j} h_i^G g_j a_j + h_i^G n$. The modification of $w_i$ is calculated as,

$$\sigma_{w_i}^2 = \frac{E_s}{N_t} (\beta_i - \beta_i^2) \tag{19}$$

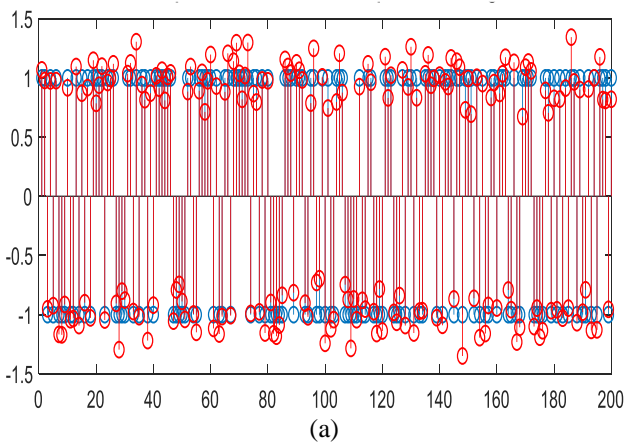

(a)

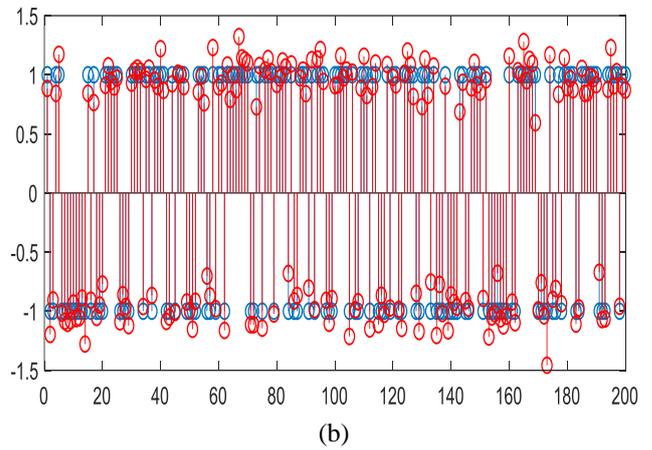

(b)

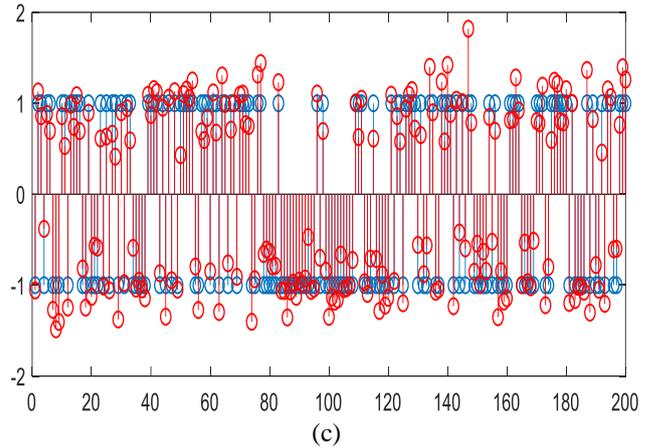

(c)

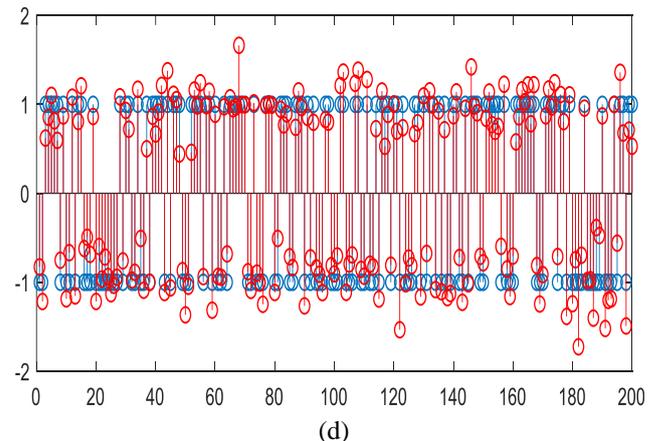

(d)

**Figure 6.** MIMO System LMMSE (a) First LE sequence for real part (b) First LE sequence for imaginary part (c) Second LE sequence for real part (d) Second LE sequence for imaginary part

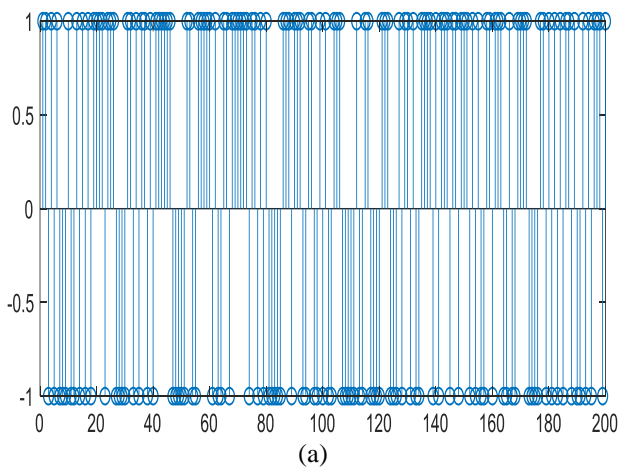

(a)

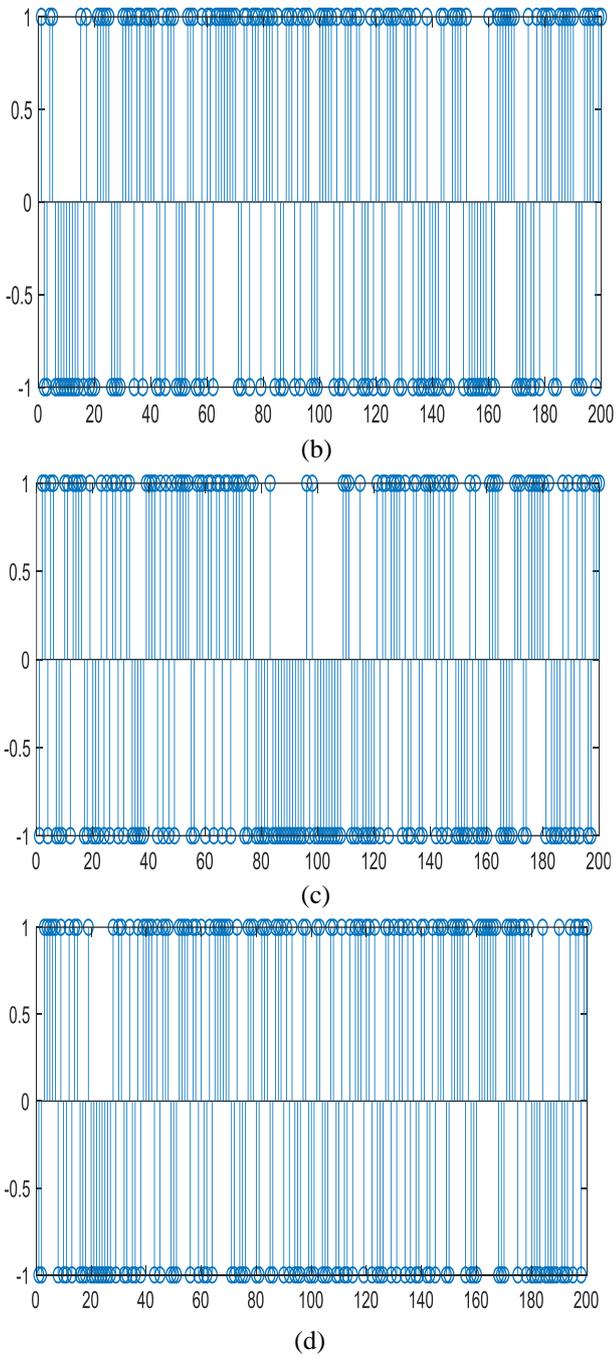

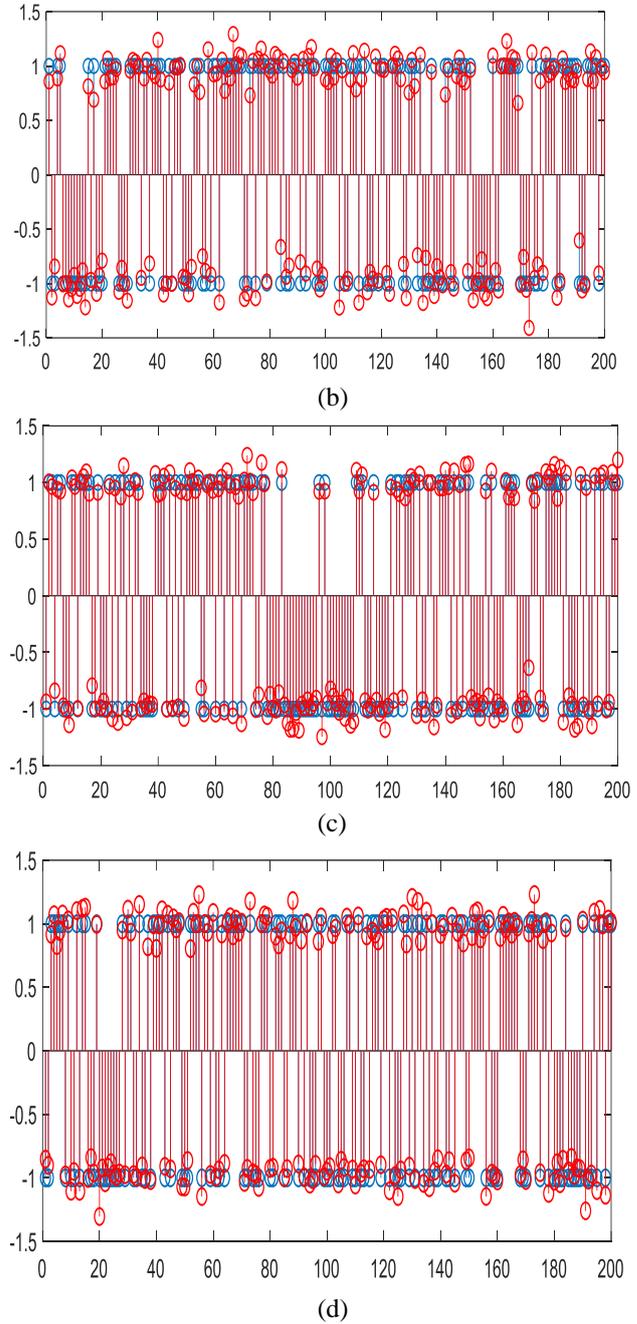

**Figure 7.** MIMO System detected sequence of LMMSE (a) First deceted LE sequence for real part (b) First deceted LE sequence for imaginary part (c) Second deceted LE sequence for real part (d) Second deceted LE sequence for imaginary part

**Figure 8.** MIMO System LMMSE (a) First MMSE sequence for real part (b) First MMSE sequence for imaginary part (c) Second MMSE sequence for real part (d) Second MMSE sequence for imaginary part (calculated by $P_d$)

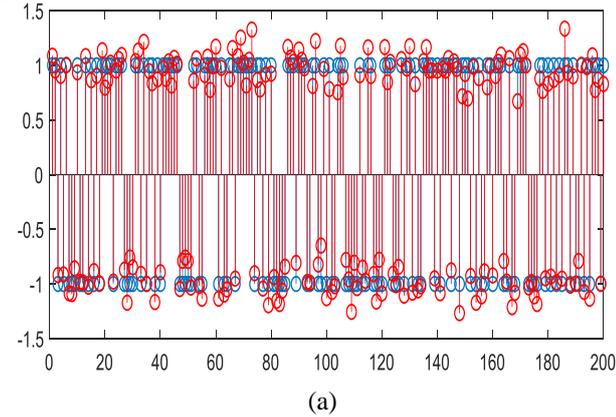

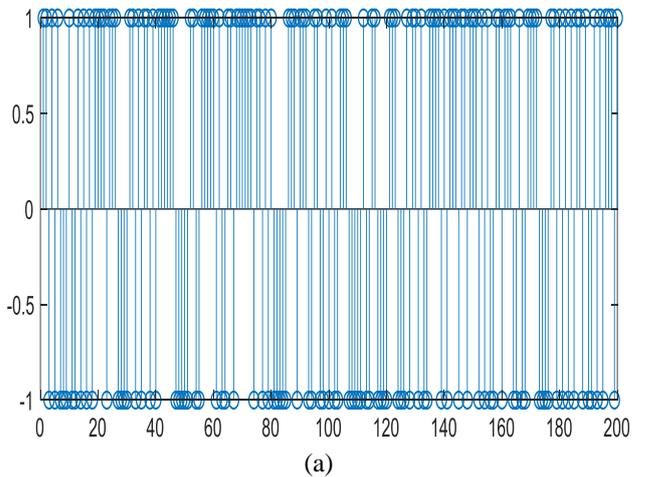

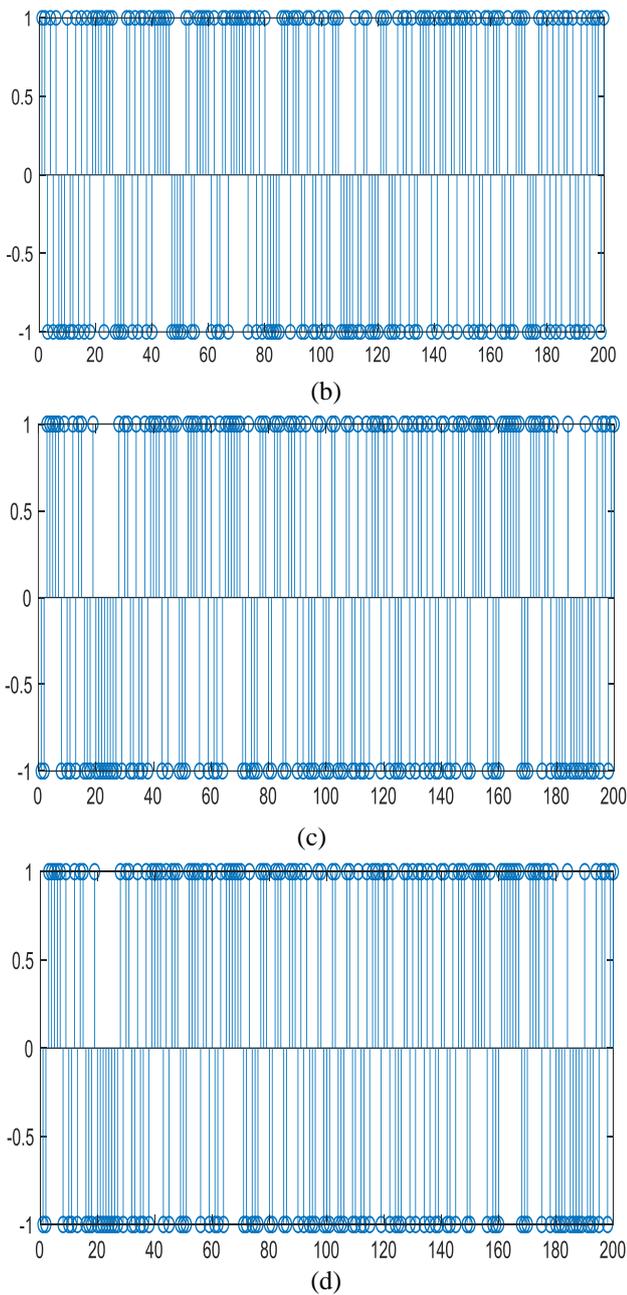

(b)

(c)

(d)

**Figure 9.** MIMO System detected sequence of LMMSE (a) First deceted MMSE sequence for real part (b) First deceted MMSE sequence for imaginary part (c) Second deceted MMSE sequence for real part (d) Second deceted MMSE sequence for imaginary part (calculated by $P_d$)

### E. Substantiation of Spatial Multiplexing Utilizing Decoupling of MIMO Scheme

For changing time and selective frequency MIMO channels using $N_t$ and $N_r$, the scheme I/O correlation can be illustrated by

$$B(t) = \int G_c(t, f) A(f) \exp(i2\pi f t) \, df + n \quad (20)$$

Where, the of signal vector output is $B(t)$ with $N_r \times 1$ dimension, the vector of input signal is $A(t)$ thru $N_t \times 1$ dimension, the vector of noise is $n$ with $N_r \times 1$ dimension and the transfer function of continuous time channel is $G_c$ with $N_t \times N_r$ dimension.

By changing the continuous time thru discrete samples then conveying the frequency selectivity with respect to discrete multipath intervals, the wideband in discrete-time MIMO channel can be illustrated via a four-dimensional channel, G transfer matrix by means of $N_q \times N_t \times N_r \times N_s$ dimensions. Wherever *Ns* and *Nq* are the channels time samples and components of resolved multipath number, respectively. The key mission of the model of MIMO channel is to build G on the word of the preferred statistical features [28].

## IV. Power Distribution for Water Fill Algorithm

The amount of subcarrier is calculated by the $P_t$ or $P_n$. dimensions

(i)While we identify $P_t$ by way of an *L* times element vector, here are subcarriers (*L*) with dissimilar overall powers. If we require $P_n$ as element vector *(N)*, this vector of noise power is the equivalent for total subcarriers. If we agree $P_n$ as a matrix in *L*-by-*N*, respectively row relates to the consistent subcarrier.

(ii)While we require scalar ($P_t$), $P_n$ defines the amount of subcarrier. If we agree $P_n$ by means of a vector of element (N), correspondingly element is the channel noise power besides there is single subcarrier. If we agree $P_n$ as a matrix *L*-by-*N*, there are subcarriers (*L*) totally consuming similar transfered power [29].

The complete volume on water filled (power distributed) is proportionate to the channel SNR [30].

Power distributed by the separate channel is known by the equation (21) as exposed in the respective method

$$Power\ Distributed = \frac{P_{transmitters} \sum_{i=1}^{n} \frac{1}{G_i}}{\sum Channels} - \frac{1}{G_i} \quad (21)$$

Where, the power of the MIMO scheme is $P_t$ which is distributed between the dissimilar channels and G is a channel matrix of the scheme. The MIMO scheme capacity is the algebraic sum of total channels capacity as well as that is assumed by the equation.

$$Capacity = \sum_{i=1}^{n} log_2(1 + Power\ Distributed \times G) \quad (22)$$

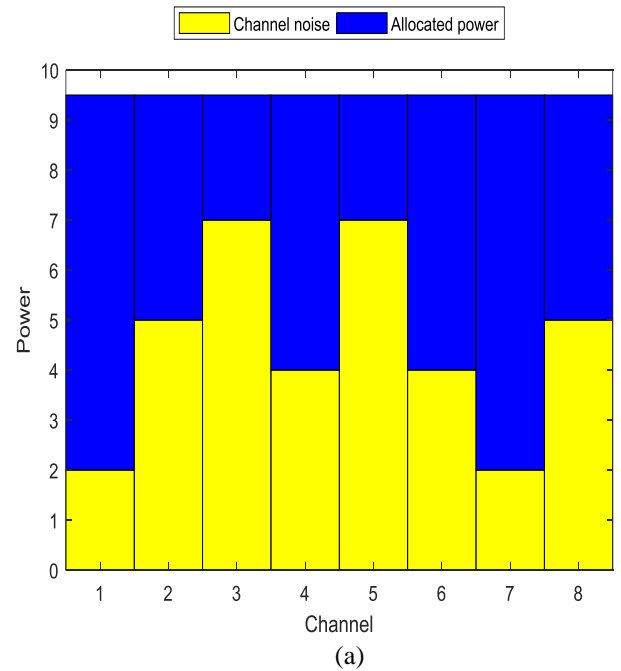

(a)

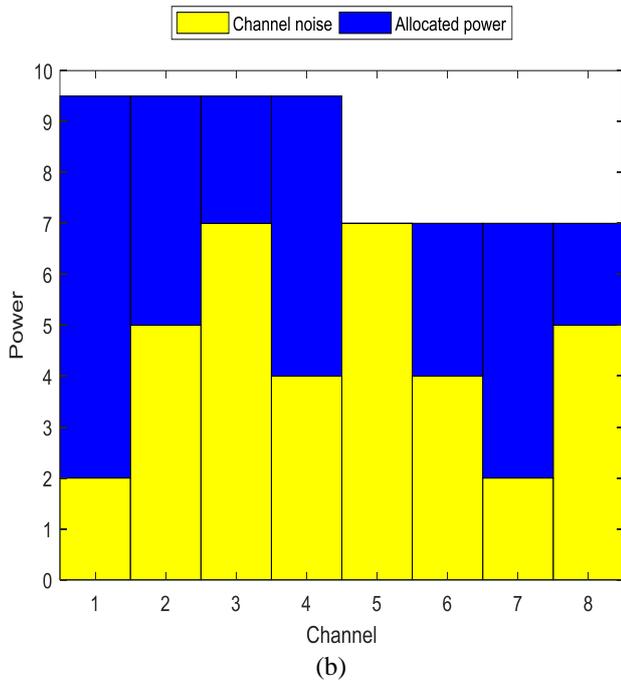

**Figure 10.** Calculation of the power distribution per channel according to two subcarriers (Similar Whole Power aimed at respective subcarrier) where (a) Pt = 20; Pn = [2 5 7 4; 7 4 2 5], (b) Pt = [20 10]; Pn = [2 5 7 4; 7 4 2 5]

Now, subsequently identifying a dissimilar entire power for respective subcarrier: $P_t = [20\ 10]$; $P_n = [2\ 5\ 7\ 4;\ 7\ 4\ 2\ 5]$; (in figure 10)

From Fig. 10(b), it can be described that distributed power per channel = $\begin{bmatrix} 7.5000 & 4.5000 & 2.5000 & 5.5000 \\ 0 & 3.0000 & 5.0000 & 2.0000 \end{bmatrix}$

---

### Algorithm 2: Waterfill Algorithm

**Input:**
(i) Enter the total power=$P_T$
(ii) Number of channels=$N_C$
(iii) Noise included in the channel=N

**Output:**
Distributed or Allocated Power Per Channel, $P_D$

    for p=1: $N_C$
        Enter the noise value as an input for the channel
end
$$[N_p \quad index] = sort\ (N);$$
$$sum(N)$$
$$for\ q = length\left(N_p\right): -1:1$$
$$r = \frac{P_T + sum\left(N_p(1:q)\right)}{q};$$
$$s = r - N_p;$$
$$P = s(1:q);$$
$$if(P(:) \geq 0);$$
$$break;$$
    end
end
$$P_D = zeros(1, N_C);$$
$$P_D(index(1:q)) = P;$$
$$Capacity = sum(\log_2(1 + P_D/N));$$

---

## V. Limited Capacity for MIMO Channel

### Algorithm 3: Limited capacity for MIMO

**Input:**
(a) Number of Realizations = n
(b)Number of transmitting antennas
(c)Number of receiving antennas
(d)Range of SNR in decibel scale
(e)Level of impairments in transmitter hardware i,e.kappa

**Output:** The average capacity over different deterministic channels, either generated synthetically with independent of Complex Gaussian Noise or taken from the measurements.

**Step 1:** Generate channel realizations based on uncorrelated Rayleigh fading
Step 2: Evaluate capacity of Ideal Hardware and capacity of Impairments for Synthetic channels
Step 3: Implement for all channel realizations
Step 3(a): for k = 1:n
    (i) Find out the Squared singular values of current channel realization
    (ii) Calculate Inversion of the squared singular values
Step 3(b): for m = 1: length (SNR)
    (i) Compute allocated power capacity for deterministic channels with the help of waterfilling algorithm
    (ii) Compute capacity with ideal hardware and a deterministic channel realization, using allocated power capacity for deterministic channels
Step 3(c): for l = 1: length (kappa)
    Determine capacity with transceiver hardware impairments for unit value of alpha with the condition of deterministic channel
end

end

end

Step 4: Calculate the limited high Signal to noise ratio irrespective of the channel distributions
Capacity Limits = A*log2(1+Number of Transmitted Antenna/(A*(kappa)²)
 where, A is the minimum value of Number of transmitting and receiving antennas

---

### A. Generalized Channel Model

Study a MIMO channel of flat-fading with $N_t$ antennas besides $N_{receive}$ antennas. The signal is received $b \in \mathbb{C}^{N_r}$ in the traditional affine established model of a channel is

$$b = \sqrt{SNR}Ga + n$$

Wherever SNR is signal to noise ratio, $a \in \mathbb{C}^{N_t}$ is the proposed signal, besides $n \sim \mathcal{CN}(0, I)$ is composite Gaussian noise circular-symmetric. $G \in \mathbb{C}^{N_t \times N_r}$ is the channel matrix supposed to be a arbitrary variable $g$.

A general MIMO channel offered and proved by capacities [11]. The mutual effect of the hardware impairments of transmitter is displayed by the deformation in transmitter $\eta_t \in \mathbb{C}^{N_t}$ and comprehensive to

$$b = \sqrt{SNR}G(a + \eta_t) + n \tag{23}$$

Where $\eta_t$ is the disparity between the proposed signal then the signal truly released by the spreader.

### B. Channel Capacity Analysis

The transmitter distinguishes the distribution of channel $f_G$ whereas the receiver distinguishes the understanding G [31]. The capacity is,

$$C_{N_t,N_r}(SNR) = \sup_{fa:t_r[E\{aa^G\}]=t_r(Q)=1} I(a;b|G) \tag{24}$$

The capacity $\mathbb{C}_{N_t,N_r}(SNR)$ can be expressed as

$$\sup_{Q:tr(Q)=1} E_G\{log_2 \det(I + SNRGQG^H(SNRG\gamma_t G^H + I)^{-1})\} \tag{25}$$

### C. Multiplexing Gains

The capacity of MIMO with ideal transceiver performs as $Mlog_2(SNR) + O(1)$, consequently it develops unlimited in the high SNR system then linearly in scales using the supposed multiplexing gain $M = \min(N_t, N_r)$ [32]. In contrast, this theorem displays that the physical MIMO capacity in channels has a limited superior bound, benevolent a dissimilar multiplexing gain:

$$M_\infty^{classic} = \lim_{SNR\to\infty} \frac{C_{N_t,N_r}(SNR)}{log_2(SNR)} = 0 \tag{26}$$

## VI. Simulation Results

Figure 11 shows the MIMO Channel average capacity for dissimilar levels of impairments in transceiver where the number of transmitting and receiving antennas is 7 and 5 respectively. The number of realization is 20000. From the figure it can be concluded that for Transmitter hardware impairments level (Value of kappa) = [0.02 0.4] capacity limits becomes [58.87, 16.43] and for Transmitter hardware impairments level, Kappa= [0.005 0.01] capacity limits becomes [78.87, 68.87].

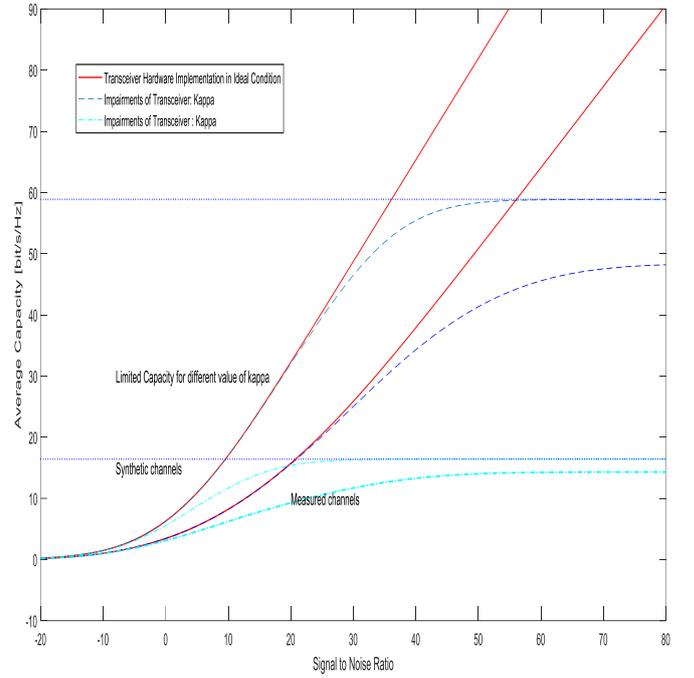

(a)

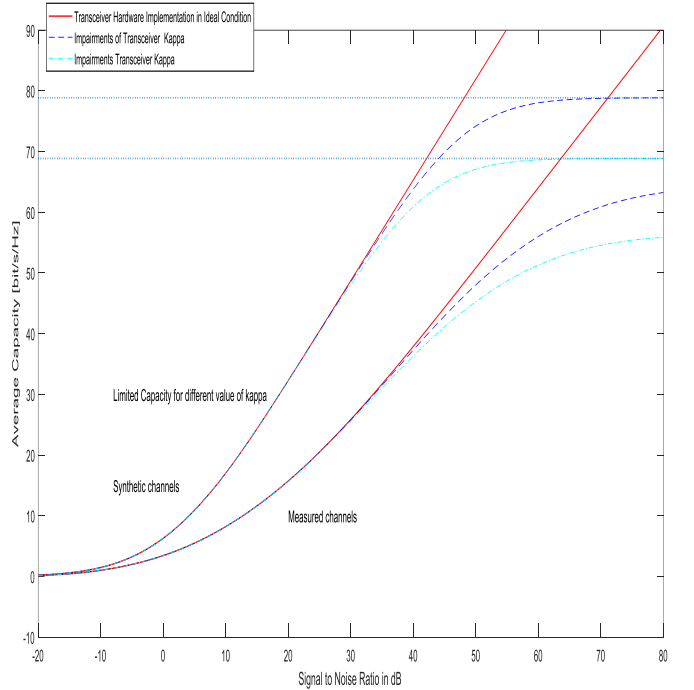

(b)

**Figure 11.** MIMO Channel average capacity for dissimilar levels of impairments in transceiver (a) Transmitter hardware impairments level, Kappa = [0.02 0.4] (b) Transmitter hardware impairments level, Kappa= [0.005 0.01]

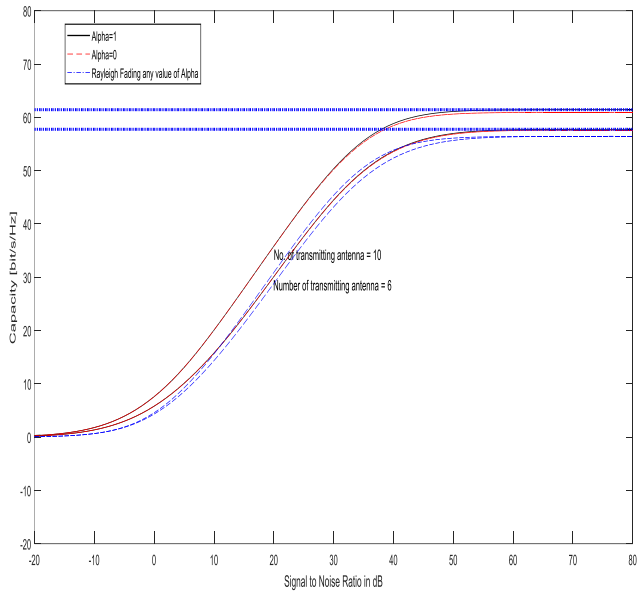

(a)

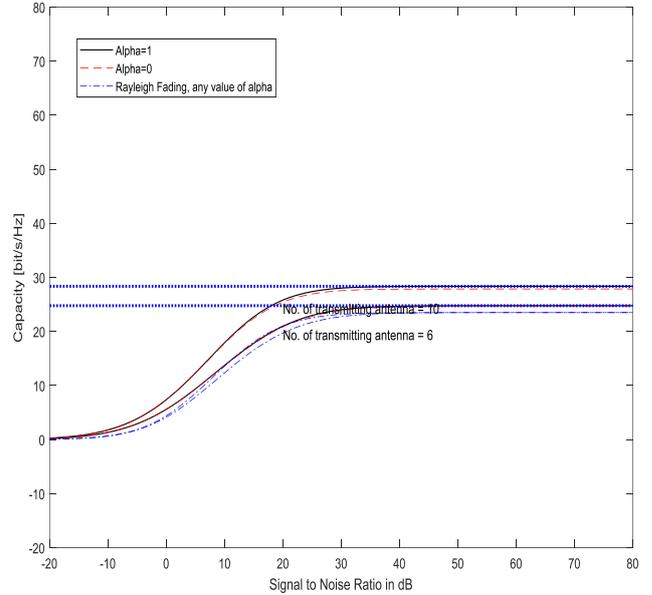

(d)

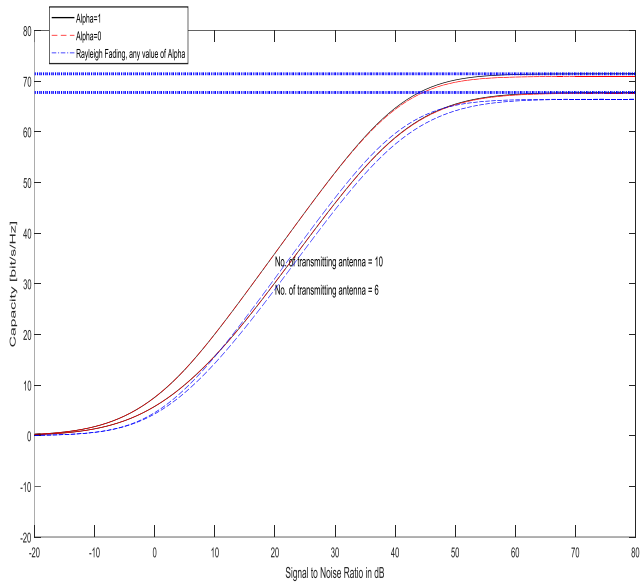

(b)

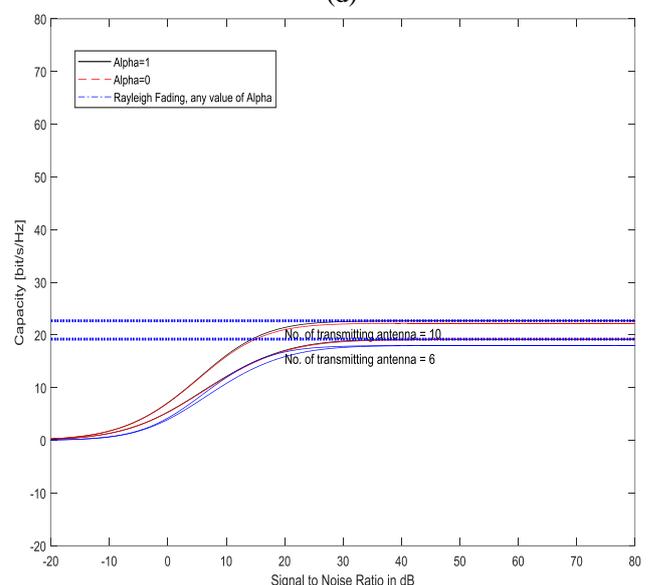

(e)

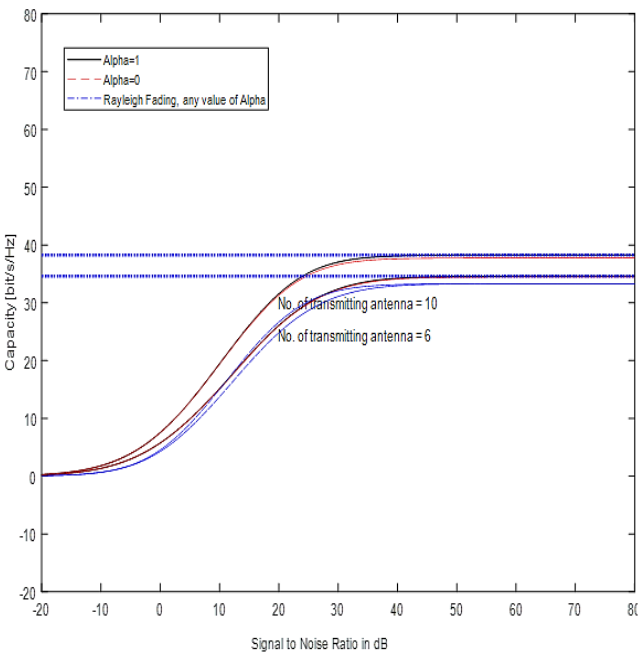

(c)

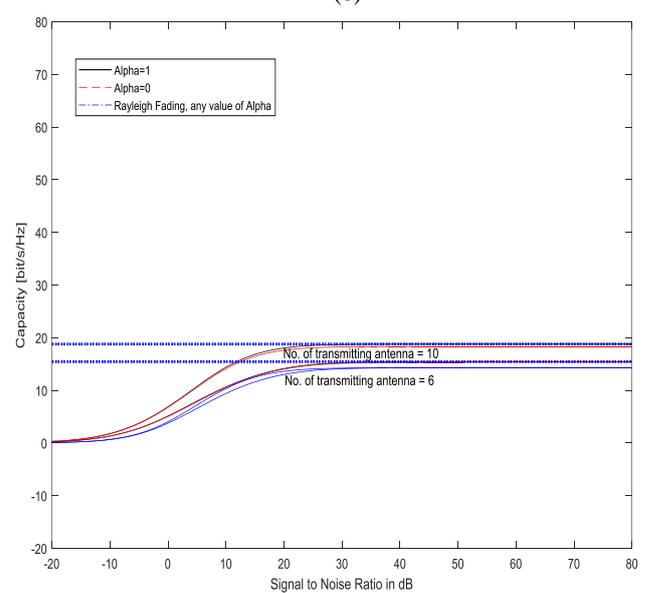

(f)

**Figure 12.** MIMO Channel capacity with number of receiving antenna= 5 and consider number of transmitting antenna in allocated channel where (a) Kappa = 0.02, (b) Kappa= 0.01,

(c) Kappa= 0.1 (d) Kappa= 0.2, (e) Kappa= 0.3 and (f) Kappa= 0.4

| Transmitter hardware impairments level (Value of Kappa) | Capacity Limits |
|---|---|
| 0.02 | [57.7561, 61.4400] |
| 0.01 | [67.7543, 71.4389] |
| 0.1 | [34.5943, 38.2553] |
| 0.2 | [24.7710, 28.3621] |
| 0.3 | [19.2065, 22.6872] |
| 0.4 | [15.4373, 18.7744] |

*Table 2.* Capacity limits for different value of Kappa

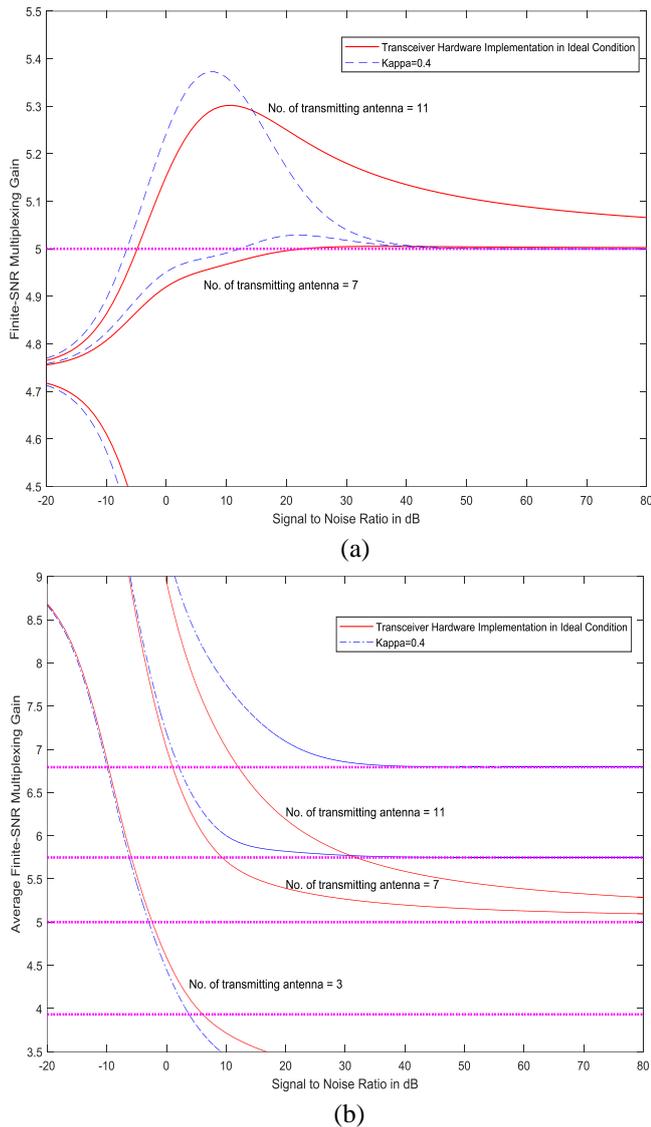

**Figure 13.** (a) Limited SNR multiplexing gain (b) Rayleigh fading channel average limited SNR multiplexing gain with transmitter hardware impairments level (Value of Kappa=0.4 for high SNR limit 6.7926 for hardware implementation for ideal condition)

| Transmitter hardware impairments level (Value of Kappa) | High SNR Limit for Ideal Hardware |
|---|---|
| 0.4 | 6.7926 |
| 0.3 | 6.4882 |
| 0.2 | 6.1775 |
| 0.1 | 5.8483 |
| 0.01 | 5.4280 |
| 0.02 | 5.5037 |

*Table 3.* Described ideal hardware limit of high SNR value for different value of Kappa

## VII. Conclusions

MIMO system utilizes antenna diversity or spatial diversity coding scheme in wireless communication system in order to mitigate the multipath fading problems. When the number of transmitting antenna ($N_t$) is equal to the number of receiving antenna ($N_r$) (i.e. 7) then average capacity is 51.0784 and 25.1878 [bit/s/Hz] respectively for Kappa (impairments level in transmitter hardware) of 0.08 and 0.3 respectively. But average capacity becomes 79.0180 and 46.6075 [bit/s/Hz] respectively for Kappa (impairments level in transmitter hardware) of 0.02 and 0.1 respectively for the same number of transmitting and receiving antenna. The MIMO Channel capacity is comparatively larger with $N_r$= 5 and $N_t$ = [6 10] in the allocated channel for Kappa of 0.01. The finite capacity limit is independent of the channel distribution. This fundamental result is elucidated by the distortion from transceiver impairments. For this reason, its power is proportional to the signal power. So, the classic multiplexing gain has become zero. The MIMO capacity grows roughly linearly with the multiplexing gain, M= min ($N_t$, $N_r$) over the SNR range. Thus provides an encouraging result and physical systems can accomplish great gains by utilizing MIMO and spatial multiplexing. Finally, it can be noted that the finite-SNR multiplexing gain decreases when extra constraints are added and so that the impairments can limit the asymptotic accuracy of channel acquisition schemes.

Efficiency Estimation for Multi Antenna MIMO User," 8[th] World Congress on Information and Communication Technologies (WICT), 2019.

## Author Biographies

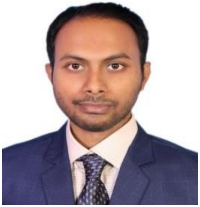

**Subrato Bharati** received his Bachelor of Science degree in Electrical and Electronic Engineering in Ranada Prasad Shaha University, Narayanganj-1400, Bangladesh. He is currently working as a research assistant at the Institute of Information and Communication Technology, Bangladesh University of Engineering and Technology, Dhaka, Bangladesh. He is a regular reviewer of ISA Transactions, Elsevier; Array, Elsevier; Vehicular Communications, Elsevier; Journal of Systems Architecture, Elsevier; Cognitive Systems Research, Elsevier; Soft Computing, Springer; Data in Brief, Elsevier, Wireless Personal Communications, Springer; Informatics in Medicine Unlocked, Elsevier. He is the guest editor of Special Issue on Development of Advanced Wireless Communications, Networks and Sensors in American Journal of Networks and Communications. His research interest includes bioinformatics, medical image processing, pattern recognition, deep learning, wireless communications, data analytics, machine learning, neural networks, distributed sensor networks, parallel and distributed computing computer networking, digital signal processing, telecommunication and feature selection. He published several IEEE, Springer reputed conference papers and also published several journals paper, Springer Book chapters.

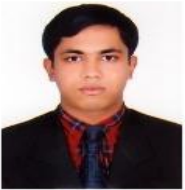

**Prajoy Podder** worked as a Lecturer in the Department of Electrical and Electronic Engineering in Ranada Prasad Shaha University, Narayanganj-1400, Bangladesh. He completed B.Sc. (Engg.) degree in Electronics and Communication Engineering from Khulna University of Engineering & Technology, Khulna-9203, Bangladesh. He is currently pursuing M.Sc. (Engg.) degree in Bangladesh University of Engineering and Technology, Institute of Information and Communication Technology, Dhaka-1000, Bangladesh. He is a researcher in the Institute of Information and Communication Technology, Bangladesh University of Engineering & Technology, Dhaka-1000, Bangladesh. He is regular reviewer of Data in Brief, Elsevier and Frontiers of Information Technology and Electronic Engineering, Springer, ARRAY, Elsevier. He is the lead guest editor of Special Issue on Development of Advanced Wireless Communications, Networks and Sensors in American Journal of Networks and Communications. His research interest includes machine learning, pattern recognition, neural networks, computer networking, distributed sensor networks, parallel and distributed computing, VLSI system design, image processing, embedded system design, data analytics. He published several IEEE conference papers, journals and Springer Book Chapters.

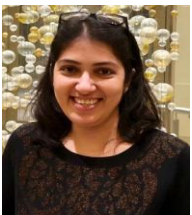

**Niketa Gandhi** is a Post-Doctoral Fellow at Department of Life Sciences, University of Mumbai working in the field of Sleep Science. Her research focuses on effects of use of social media on the sleep quality of the population before going to bed in the night. She is a Senior Member IEEE and has 15+ years of experience in academics performing various roles in teaching, administration, and research at college and university level.

Dr. Gandhi received her Ph.D. in Computer Science from the University of Mumbai and she has her Bachelor and Master of Science along with Master of Philosophy in Computer Science as well from India.

In the past Dr. Gandhi has served various positions such as faculty at Cambridge International Academy, Canada, Technical Manager at Machine Intelligence Research Labs (MIR Labs), USA, assistant professor on contract at the Department of Computer Science, University of Mumbai, India, Coordinator for Bachelors and Masters of Information Technology course at the Institute of Distance and Open Learning (IDOL), University of Mumbai and a permanent faculty position at an affiliated college of University of Mumbai.

Dr. Gandhi is a very prolific researcher who has gained a stellar reputation in academic pursuits and is evident from the extensive list of her publications.

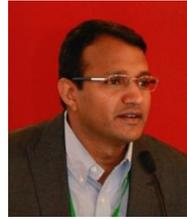

**Dr. Abraham** is the Director of Machine Intelligence Research Labs (MIR Labs), a Not-for-Profit Scientific Network for Innovation and Research Excellence connecting Industry and Academia. The Network with Headquarters in Seattle, USA has currently more than 1,000 scientific members from over 100 countries. As an Investigator / Co-Investigator, he has won research grants worth over 100+ Million US$ from Australia, USA, EU, Italy, Czech Republic, France, Malaysia and China. Dr. Abraham works in a multi-disciplinary environment involving machine intelligence, cyber-physical systems, Internet of things, network security, sensor networks, Web intelligence, Web services, data mining and applied to various real world problems. In these areas he has authored / coauthored more than 1,300+ research publications out of which there are 100+ books covering various aspects of Computer Science. One of his books was translated to Japanese and few other articles were translated to Russian and Chinese. About 1000+ publications are indexed by Scopus and over 900 are indexed by Thomson ISI Web of Science. Some of the articles are available in the ScienceDirect Top 25 hottest articles. He has 700+ co-authors originating from 40+ countries. Dr. Abraham has more than 37,000+ academic citations (h-index of 90 as per google scholar). He has given more than 100 plenary lectures and conference tutorials (in 20+ countries). For his research, he has won seven best paper awards at prestigious International conferences held in Belgium, Canada Bahrain, Czech Republic, China and India. Since 2008, Dr. Abraham is the Chair of IEEE Systems Man and Cybernetics Society Technical Committee on Soft Computing (which has over 200+ members) and served as a Distinguished Lecturer of IEEE Computer Society representing Europe (2011-2013). Currently Dr. Abraham is the editor-in-chief of Engineering Applications of Artificial Intelligence (EAAI) and serves/served the editorial board of over 15 International Journals indexed by Thomson ISI. He is actively involved in the organization of several academic conferences, and some of them are now annual events. Dr. Abraham received Ph.D. degree in Computer Science from Monash University, Melbourne, Australia (2001) and a Master of Science Degree from Nanyang Technological University, Singapore (1998). More information at: http://www.softcomputing.net/